\documentclass[showpacs,amssymb,preprint,preprintnumbers,nofootinbib,superscriptaddress]{revtex4}
\usepackage{amsmath}
\usepackage{graphicx}
\usepackage{latexsym}
\usepackage{amsfonts}
\usepackage{url,hyperref}
\usepackage{bm}
\usepackage{textcomp}
\usepackage{color}
\newcommand{\be}{\begin{equation}}
\newcommand{\ee}{\end{equation}}
\newcommand{\bea}{\begin{eqnarray}}
\newcommand{\eea}{\end{eqnarray}}
\newcommand{\Rlabcd}{R_{abcd}}
\newcommand{\Ruabcd}{R^{abcd}}
\newcommand{\tg}{\tilde{g}}
\newcommand{\tR}{\tilde{R}}

\newcommand{\om}{\omega}

\newcommand{\tH}{\tilde{H}}

\begin{document}
\title{Conformally-related Einstein-Langevin equations \\ for metric fluctuations in
stochastic gravity}
\author{Seema Satin}
\email[Email: ]{sesatin@tku.edu.tw}
\affiliation{Department of Physics, Tamkang University, Tamsui, Taipei, Taiwan}
\author{H.~T.~Cho}
\email[Email: ]{htcho@mail.tku.edu.tw}
\affiliation{Department of Physics, Tamkang University, Tamsui, Taipei, Taiwan}
\author{Bei Lok Hu}
\email[Email: ]{blhu@umd.edu}
\affiliation{Maryland Center for Fundamental Physics and Joint Quantum Institute, University of Maryland, College Park, Maryland 20742-4111, USA}


\vspace{10pt}

\begin{abstract}
For a conformally-coupled scalar field we obtain the conformally-related Einstein-Langevin equations, using appropriate transformations for all the quantities in the equations between two conformally-related spacetimes. In particular, we analyze the transformations of the influence action, the stress energy tensor, the noise kernel and the dissipation kernel. In due course the fluctuation-dissipation relation is also discussed. The analysis in this paper thereby facilitates a general solution to the Einstein-Langevin equation once the solution of the equation in a simpler, conformally-related spacetime is known. For example, from the Minkowski solution of Martin and Verdaguer, those of the Einstein-Langevin equations in conformally-flat spacetimes, especially for spatially-flat Friedmann-Robertson-Walker models, can be readily obtained.
\end{abstract}

\maketitle

\section{Introduction}

Quantum field theory in curved spacetime developed in the '70s (e.g., \cite{BirDav,ParTom}) and ensuing studies of the \textit{effects of quantum matter fields} on spacetime dynamics  (semiclassical gravity theory SCG)  (e.g., \cite{FHH79,HH79,CH87,CV94}) have provided the framework for important theoretical predictions, Hawking radiation and the inflationary cosmology being two well known examples.   Inquires into the nature and role played by \textit{the fluctuations of quantum fields} began in the '80s (e.g., \cite{Ford,FFR}).  For the description of the effects (backreaction) of quantum field fluctuations on the spacetime dynamics,  the theory of semiclassical stochastic gravity \cite{HV}  was developed systematically in the '90s. The central question asked there is, specifically,  how metric fluctuations  (note,  \textit{fluctuation} is a stochastic function, not \textit{perturbation,} a deterministic function)  behave through the solutions of the Einstein-Langevin equation (ELE)\cite{ELE} with the  correlation functions of the quantum stress tensor as source.  ELE is an upgrade of the semiclassical Einstein equation (SCE) with the expectation values of the stress energy tensor as source to include the (quantum) matter field and (classical) gravitational field fluctuation effects.    Stochastic gravity theory has been used to provide certain criteria  for the validity of SCG \cite{HRV}, and applied to problems of interest in early universe \cite{rouver,enric} and black hole physics  \cite{SRH,HuRou}.

Using stochastic gravity theory to investigate the behavior of the metric fluctuations involves three stages of work: 1) Computation of the noise kernel, defined as the expectation value of the stress energy bitensor, for the quantum field matter source in the class of curved spacetime of interest; 2) Derivation of the ELE for the class of spacetimes of interest,  with the background spacetime obtained from the solutions of the SCE; 3) Solutions of the ELE.  A clear demonstratation of these procedures for metric fluctuations in Minkowski spacetime can be found in \cite{ver1,ver2}.

For  task 1) expressions of the noise kernel (NK) for various classes of curved spacetimes have been obtained from direct computations,  using the zeta function \cite{PH97,chohu1} or point separation \cite{phihu0} methods, or  using the general expression of the NK  expressed in terms of the higher order covariant derivatives of the Green function \cite{phihu1} under certain  approximations \cite{phihu2,EBRAH}.  There is a slow yet steady stream of activities in the last decade.  Another approach is to exploit the symmetry relations between different classes of spacetimes, such as the derivation of NKs \cite{chohu2} for spacetimes related by conformal transformations and by the thermalization relation, as  first proposed by \cite{candow}.  Our present work continues in this vein for task 2), namely, using the conformal transformations to derive new ELE from known ones,  such as the ELE in all the spacetimes which are conformally-related to Minkowski, where the ELE has been obtained earlier \cite{ver2}, and curved spacetimes possessing high degree of symmetry, where ELE is relatively easy to derive (some such spacetimes are described in \cite{chohu2}).  We shall  in future work derive the ELE for  the other class of spacetimes relatable through thermalization transformations. By then one should be  sufficiently equipped to tackle task 3) proper.

We begin with Section II to obtain the conformal transformation of the influence action. In Section III conformally-related stress tensors are obtained \cite{ottebrown} including the effect of the conformal anomaly. We also give a comparison of our results with  those of nearly-flat spacetime that have been obtained earlier by \cite{star,horo,horo1,horo2}. This shows consistency of our results for the quantum stress tensor via the influence action method with those via other methods in earlier works. Then in Section IV we lay out the transformation of the complete ELE by finding the appropriate transformations for various quantities in the equation. The expression for two dissipation kernels in conformally-related spacetime is worked out in Section V. There a fluctuation-dissipation relation for the relevant case is presented.  We draw the conclusions in Section VI.

\section{The influence action} \label{sec:action}

In the theory of stochastic gravity one considers gravity as the
system and quantum fields as the environment.
Stochastic gravity (including quantum fluctuations) is based on the ELE
\be
G^{ab} [g + h]+\Lambda(g^{ab}-h^{ab}) = 8\pi G (\langle T^{ab}[g+h]\rangle + \xi^{ab}\label{ELE}
[g])
\ee
to linear order in $h$. Here $G^{ab}$ is the Einstein tensor, $\Lambda$ is the cosmological constant, $\langle T^{ab}[g+h]\rangle$ is the in-in expectation value of quantum field stress tensor,
and $\xi^{ab}$ is the stochastic force induced by
quantum field fluctuations.
This equation can be obtained through influence functional method or the axiomatic
approach as has been shown in \cite{HV}.

For the influence functional method, the CTP effective action is
given by
\begin{eqnarray}
e^{i \Gamma[g_+,g_-]} & = & e^{i S_g[g_+]- i S_g[g_-]} \int_{CTP} D \phi_+
 D \phi_- e^{i S_m[\phi_+,g_+] - i S_m[\phi_-,g_-]} \\
 &=& e^{S_g[g_+] - i S_g[g_-] + i S_{IF} [ g_+,g_-]}
\end{eqnarray}
where $S_g$ and $S_m$ are the gravity and quantum field actions, respectively.
The influence action $S_{IF}$ is due to the quantum field, which after expansion
in terms of $g_\pm= g+ h_\pm$ gives
\begin{eqnarray}
S_{IF} & = & \frac{1}{2} \int d^4x \sqrt{-g(x)} \langle T^{ab}(x) \rangle
\Delta h_{ab} (x) \nonumber \\
& & - \frac{1}{8} \int d^4x\, d^4y \sqrt{-g(x)} \sqrt{-g(y)} \Delta
h_{ab}(x) 
[ K^{abcd}(x,y) + H^{abcd}(x,y)] \Sigma h_{cd}(y)  \nonumber \\
& & + \frac{i}{8} \int d^4x\, d^4y \sqrt{-g(x)} \sqrt{-g(y)} \Delta h_{ab}(x)
N^{abcd}(x,y) \Delta h_{cd}(y) \label{IF}
\end{eqnarray}
where $\Delta h_{ab}\equiv h_{+ab}-h_{-ab}$ and $\Sigma h_{ab}\equiv h_{+ab}+h_{-ab}$.
The detailed forms of the kernels $N^{abcd}(x,y)$, $H^{abcd}(x,y)$, and $K^{abcd}(x,y)$ can be found in \cite{HV}.
$N^{abcd}(x,y)$ is the so-called noise kernel while the antisymmetric part of $H^{abcd}(x,y)$ is related to the dissipation kernel.

It is of interest to seek the conformal transformations between ELEs of conformally-related spacetimes.
This would play a very signifincant role in determining solutions of this equation from known ones.
We begin by considering the conformal transformation of the influence action.
This proceeds as follows. Under a conformal transformation,
\begin{equation}
\tilde{g}_{ab} = e^{-2 \omega} g_{ab}\ \  ;\ \  \tilde{\phi} = e^\omega \phi ,
\end{equation}
the quantum field action $S_{m}$ will not change if $\phi$ is a conformally-coupled scalar field.
However,
due to the conformal anomaly \cite{FHH79,ottebrown,ottepage},
\bea
\int D \tilde{\phi}\, e^{i S_m[\tilde{\phi},\tilde{g}]}
&=& \int D \phi \,e^{i(S_{m}[\phi,g]-aP[g,\omega]-bQ[g,\omega])}
\eea
where $a=1/1920\pi^2$, $b=-1/5760\pi^2$, and
\bea
 P[g,\omega] &= & \int d^4x \sqrt{-g}\,\Big\{(\Rlabcd \Ruabcd- 2 R_{ab}R^{ab}
+\frac{1}{3} R^2) \omega \nonumber \\ \nonumber
& & \ \ \ \ \ \ \ \ \ \ \ \ \ \ \ \ \ +
\frac{2}{3} [ R + 3 (\Box\omega-\omega_{;a}
\omega^{;a})](\Box\omega - \omega_{;b}\omega^{;b})\Big\}\\
Q[g,\omega] & = & \int d^4x \sqrt{-g}\, \Big\{(\Rlabcd \Ruabcd - 4 R_{ab}R^{ab} +R^2)
\omega + 4 R_{ab} \omega^{;a}\omega^{;b} \nonumber\\
& & \ \ \ \ \ \ \ \ \ \ \ \ \ \ \ \ \ - 2 R\, \omega_{;a} \omega^{;a} + 2 (\omega_{;a}\omega^{;a})^2 -
4 \omega_{;a} \omega^{;a} \Box \omega \Big\}
\eea

Hence the transformation of the influence action can be seen to  be of
the form,
\be
S_{IF} [ \tilde{g}_+,\tilde{g}_-] = S_{IF}[g_+,g_-] -a P[g_+,\omega_+] -
b\, Q[g_+,\omega_+]+a P[g_-, \omega_-] +b\, Q[g_-,\omega_-]\label{coninfact}
\ee
As from the results in \cite{ottebrown}, $S_{IF}$ and the further stress tensor obtained from it are renormalized.  Note the effective actions in \cite{ottebrown}, as in most earlier work,  are of the in-out type (Schwinger-DeWitt).  What we have here is the in-in (Schwinger-Keldysh) version,  which is the physically correct quantity to use for the derivation of real and causal equations of motion \cite{CH87}.

\section{The stress energy tensor} \label{sec:stress}

The renormalized expectation value of the stress tensor is given by
\be
\langle T^{ab} (x)\rangle = \left.\frac{2}{\sqrt{-g(x)} } \frac{ \delta S_{IF}}{\delta
g_{+ab}(x)}\right|_{g_+ = g_-}
\ee
Taking the variation of the influence action with respect to $g_{+ab}$ of Eq.~(\ref{coninfact}),
\be  \label{eq:mat}
\frac{\delta S_{IF}[\tilde{g}_+,\tilde{g}_-]}{\delta{\tilde{g}_{+ab}}}
 = e^{2 \omega} \frac{\delta}{\delta g_{+ab}}( S_{IF}[g_+,g_-]
- a P[g_+,\omega_+] - b Q[g_+,\omega_+])
\ee
The variations of $P[g,\omega]$ and $Q[g,\omega]$ are given in \cite{ottepage}. Denote
\begin{eqnarray}
E^{ab} &=& \frac{2}{\sqrt{-g}} \frac{\delta P}{\delta g_{ab}}\nonumber\\
&=&4 e^{-6 \omega}[2
(\tilde{C}^{cabd}\omega)_{;cd} + \tilde{R}_{cd} \tilde{C}^{cabd}\omega]
+ \frac{1}{9}[ e^{-6 \omega} \tilde{B}^{ab} - B^{ab}]\label{defeab}
\end{eqnarray}
and
\begin{eqnarray}
F^{ab} &=& \frac{2}{\sqrt{-g}} \frac{\delta Q}{\delta g_{ab}}\nonumber\\
&=&e^{-6 \omega} ( 4
\tilde{C}^{cabd} \tilde{R}_{cd} + 2 \tilde{H}^{ab}) - ( 4 C^{cabd} R_{cd}
+ 2 H^{ab})\label{deffab}
\end{eqnarray}
where
\bea
B^{ab} & = & \frac{1}{\sqrt{-g}} \frac{\delta}{\delta g_{ab}} \int
d^4 x \sqrt{-g} R^2 \nonumber \\
&=& 2 R^{;ab} - 2 R R^{ab} + (\frac{1}{2}R^2 - 2 \Box R) g^{ab} \label{tensorB}
\eea
\be
H^{ab} = - R^{ac}R_c^b + \frac{2}{3} R R^{ab} + (\frac{1}{2} R^{cd}R_{cd}
- \frac{1}{4} R^2) g^{ab}
\ee
The conformal transformation of the stress energy tensor takes the form
\be \label{eq:stress}
\langle T^{ab} [\tilde{g}] \rangle = e^{6 \omega}\left( \langle T^{ab}[g] \rangle -
a E^{ab} - b F^{ab}\right)
\ee

Using this transformation closed expressions for the stress energy tensors of conformally flat spacetimes can be obtained \cite{candow}. Since $g_{ab}$ is the flat metric, one has
\begin{eqnarray}
\langle T^{ab} [\tilde{g}] \rangle = e^{6 \omega} \langle T^{ab}[g] \rangle -\frac{a}{9}\tilde{B}^{ab}-2b\tilde{H}^{ab}
\end{eqnarray}
As long as $\langle T^{ab}[g] \rangle$ is known exactly, either in the Minkowski or the Rindler vacua, one also has an exact expression for $\langle T^{ab} [\tilde{g}] \rangle$.

Further comparison can be done for weakly perturbed conformally flat spacetimes with the results given in \cite{horo,horo1,horo2}. From Eq.~(\ref{eq:stress}) we have
\begin{eqnarray}
\langle T^{ab} [\tilde{g}+\tilde{h}] \rangle = e^{6 \omega}\left( \langle T^{ab}[\eta+h] \rangle -
a E^{ab}[\eta+h] - b F^{ab}[\eta+h]\right)
\end{eqnarray}
To first order in $h$,
\begin{equation}
\langle T^{ab} [\tilde{g}+\tilde{h}] \rangle = \langle T^{ab} [\tilde{g}] \rangle
+\langle T^{(1)ab} [\tilde{h}] \rangle
\end{equation}
From \cite{horo}, also to first order in $h$, the stress energy tensor in perturbed Minkowski spacetime is given by
\begin{equation}
\langle T^{ab}[\eta+h] \rangle = -\frac{a}{9} B^{(1)ab}[h]+a\int\,d^{4}x' H_{\lambda}(x-x')A^{(1)ab}[h]
\end{equation}
where
\bea
A^{ab} & = &   \frac{1}{\sqrt{-g}} \frac{\delta}{\delta g_{ab}}
\int d^4 x \sqrt{-g} C_{cdef} C^{cdef} \nonumber \\
& = & \frac{1}{2}g^{ab}C_{cdef}C^{cdef} -
 2 R^{acde}R^b_{\ cde} + 4 R^{ac} R_c^b - \frac{2}{3} R R^{ab} -
 2 \Box R^{ab} + 
\frac{2}{3} R^{;ab} + \frac{1}{3} g^{ab}\Box R, \nonumber\\
\eea
and
\begin{equation}
A^{(1)ab}=-2\partial^{2}R^{(1)ab}+\frac{1}{3}(2\partial^{a}\partial^{b}+\eta^{ab}\partial^{2})R^{(1)}.
\end{equation}
Similarly, the first order expansion of $B^{ab}$ in Eq.~(\ref{tensorB}) gives
\begin{equation}
B^{(1)ab}=2(\partial^{a}\partial^{b}-\eta^{ab}\partial^{2})R^{(1)}.
\end{equation}
The kernel $H_{\lambda}$ has the Fourier transform \cite{horo,ver2}
\begin{equation}
H_{\lambda}(x-x')=\int \frac{d^{4}p}{(2\pi)^4}e^{ip(x-x')}\left[i\pi({\rm sgn}\ p^{0})\theta(-p^2)-{\rm ln}\left(\lambda^2\left|p^{2}\right|\right)\right].
\end{equation}
The expansions to first order in $h$ for the tensors $E^{ab}$ and $F^{ab}$ are
\begin{eqnarray}
E^{ab}[\eta+h]&=&\frac{1}{9}e^{-6\omega}\tilde{B}^{ab}+8\partial_{c}\partial_{d}\left(C^{(1)cabd}\omega\right)
+\frac{1}{9}e^{-6\omega}\tilde{B}^{(1)ab}-\frac{1}{9}B^{(1)ab}\\
F^{ab}[\eta+h]&=&2e^{-6\omega}\tilde{H}^{ab}+e^{-6\omega}\left(4\tilde{C}^{(1)cabd}\tilde{R}_{cd}+2\tilde{H}^{(1)ab}\right)
\end{eqnarray}
where we have used the identity
\begin{eqnarray}
e^{-6 \omega}[2(\tilde{C}^{cabd}\omega)_{;cd} + \tilde{R}_{cd} \tilde{C}^{cabd}\omega]
=2(C^{cabd}\omega)_{;cd} + R_{cd} C^{cabd}\omega
\end{eqnarray}
for any two conformally-related spacetimes. Now
combining the above results we have
\begin{eqnarray}
\langle T^{(1)ab} [\tilde{h}] \rangle
&=&a\,e^{6\omega}\int\,d^{4}x' H_{\lambda}(x-x')A^{(1)ab}[h]
-8a\,e^{6\omega}\partial_{c}\partial_{d}\left(C^{(1)cabd}\omega\right)\nonumber\\
&&\ \ -\frac{a}{9}\tilde{B}^{(1)ab}
-2b\left(2\tilde{C}^{(1)cabd}\tilde{R}_{cd}+\tilde{H}^{(1)ab}\right)\label{horowald}
\end{eqnarray}
which agrees with the result in \cite{horo1,horo2}. Note that in \cite{horo1,horo2} Eq.~(\ref{horowald}) was obtained
from the general requirements on the stress energy tensor. Here we have derived the same expression in a more direct manner.

\section{The Einstein-Langevin equation} \label{sec:EL}
After considering the conformal transformation of the stress energy tensor in the last section, we shall look at the stochastic force $\xi^{ab}$, another ingredient of the ELE in Eq.~(\ref{ELE}). The two point correlation function of $\xi^{ab}(x)$ is given by
\begin{equation}
\langle \xi^{ab}(x)\xi^{cd}(y)\rangle=N^{abcd}(x,y),
\end{equation}
where the brackets denote stochastic average.
The noise kernel $N^{abcd}(x,y)$ is also the symmetric two point correlation function of the stress energy tensor. It has been the focus of various works \cite{phihu1,phihu2,chohu1,EBRAH,batcho,chohu2,seema} on the backreaction of the quantum field fluctuations onto the background spacetime. In particular, in \cite{EBRAH} the transformation between the noise kernels in two conformally-related spacetimes $g_{ab}$ and $\tilde{g}_{ab}$ are derived. It is given by
\begin{equation}
\tilde{N}^{abcd}(x,y)=e^{6\omega(x)}N^{abcd}(x,y)e^{6\omega(y)}.\label{confnoise}
\end{equation}

The gravitational part of the ELE can be derived from the gravity action. With renormalization taken into account it can be written as
\be
\tilde{S}_g [\tilde{g}] =  \int d^4 x \sqrt{-\tg} \,\left[ \frac{1}{8 \pi G}\left(\frac{1}{2} \tR -
 \Lambda\right) + \alpha\, \tilde{C}_{cdef} \tilde{C}^{cdef} + \beta \tR^2\right]
\ee
where the parameters $ G,\Lambda, \alpha $ and $\beta$ are renormalized
constants. This action gives rise to the gravity sector of the
semiclassical Einstein equation for metric $\tg_{ab}+\tilde{h}_{ab}$ which takes the form
\be
 \frac{1}{8 \pi G}(\tilde{G}^{ab} [\tilde{g}+\tilde{h}] +
\Lambda(\tilde{g}^{ab} -\tilde{h}^{ab})) - 2 ( \alpha \tilde{A}^{ab} +
 \beta \tilde{B}^{ab})[\tilde{g}+\tilde{h}]
\ee
Using the conformal transformation in Eq.~(\ref{eq:stress}) between the stress energy tensors $T^{ab}[\tilde{g}+\tilde{h}]$ and $T^{ab}[g+h]$, the corrresponding ELE for the conformally-related spacetime with metric $\tilde{g}_{ab}+\tilde{h}_{ab}$ reads
\bea
& &  \frac{1}{8 \pi G}(\tilde{G}^{ab} [\tilde{g}+\tilde{h}] +
\Lambda(\tilde{g}^{ab} - \tilde{h}^{ab})) - 2 ( \alpha \tilde{A}^{ab} +
 \beta \tilde{B}^{ab})[\tilde{g}+\tilde{h}]\nonumber \\
&= & e^{6 \omega}( T_R^{ab} -  a E^{ab}
 -b F^{ab})[g+h] + 2 \tilde{\xi^{ab}}
\eea
where $E^{ab}$ and $F^{ab}$ are given by the expressions in Eqs.~(\ref{defeab}) and (\ref{deffab}) respectively,
with ($g_{ab}, \tilde{g}_{ab}$) replaced by  ($g_{ab}+h_{ab},  \tilde{g}_{ab}+\tilde{h}_{ab}$) .

Assume that $\tilde{g}_{ab}$ satisfies the semiclassical Einstein equation
\begin{equation}
\frac{1}{8 \pi G}(\tilde{G}^{ab} [\tilde{g}] +
\Lambda\tilde{g}^{ab}) - 2 ( \alpha \tilde{A}^{ab} +
 \beta \tilde{B}^{ab})[\tilde{g}]
=  e^{6 \omega}( T_R^{ab} -  a E^{ab}
 -b F^{ab})[g].
\end{equation}
Then, to the first order in $h_{ab}$ or $\tilde{h}_{ab}=e^{-2\omega}h_{ab}$, the ELE for the conformally-related spacetime can be expressed as
\bea
 \frac{1}{8 \pi G}(\tilde{G}^{(1)ab} [\tilde{h}] -
\Lambda\tilde{h}^{ab}) - 2 ( \alpha \tilde{A}^{(1)ab} +
 \beta \tilde{B}^{(1)ab})[\tilde{h}]
=  e^{6 \omega}( T_R^{(1)ab} -  a E^{(1)ab}
 -b F^{(1)ab})[h] + 2 \tilde{\xi}^{ab}\nonumber\\ \label{pertELE}
\eea
In this form, as long as $T_{R}^{(1)ab}$ is known, this equation can be used to solve for the perturbation $\tilde{h}_{ab}$
in the background metric $\tilde{g}_{ab}$. One prominent example would be the various Friedmann-Robertson-Walker models.
One can analyze the perturbations of a conformally-coupled field of these models just by putting in the corresponding
exact quantum stress tensors in the static Minkowski, Einstein, and open Einstein spacetimes \cite{candow,chohu2}. This will be pursued in our subsequent works.

As we have discussed in the previous section, the case with spacetime conformal to Minkowski is particularly simple.
The perturbed ELE can be written as
\begin{eqnarray}
&& \frac{1}{8 \pi G}(\tilde{G}^{(1)ab} [\tilde{h}] -
\Lambda\tilde{h}^{ab}) - 2 \left[ \alpha \tilde{A}^{(1)ab} +
 \left(\beta-\frac{a}{18}\right) \tilde{B}^{(1)ab}\right][\tilde{h}]\nonumber\\
&=&a\,e^{6\omega}\int\,d^{4}x' H_{\lambda}(x-x')A^{(1)ab}[h]
-8a\,e^{6\omega}\partial_{c}\partial_{d}\left(C^{(1)cabd}\omega\right)\nonumber\\
&&\ \
-2b\left(2\tilde{C}^{(1)cabd}\tilde{R}_{cd}+\tilde{H}^{(1)ab}\right)+2\tilde{\xi}^{ab}\label{pertELEflat}
\end{eqnarray}
This equation can be readily applied to the study of perturbations on the spatially-flat Friedmann universes useful for the problem of structure formation  
in inflationary cosmology.  As Roura and Verdaguer \cite{rouver} have shown,  the stochastic gravity theory produces results equivalent to those obtained by quantizing the linear metric perturbations in the conventional approaches (and more notably, what the conventional approaches cannot, such as the quantized second order perturbations).

\section{The fluctuation-dissipation relation}
In a general open system the noise and the dissipation kernels are related by the so-called fluctuation-dissipation relation. For the ELE in Eq.~(\ref{pertELE}) the noise kernel is the correlation function of the stochastic force $\xi^{ab}$, while the dissipation is included in the non-local part of $T_{R}^{(1)ab}$. In \cite{ver1} the dissipation kernel was identified as the antisymmetric part of the kernel $H^{abcd}$ in Eq.~(\ref{IF}), that is,
\begin{eqnarray}
T_{R}^{(1)ab}[h]\sim \int d^{4}y\sqrt{-g(y)}\,H_{A}^{abcd}(x,y)h_{cd}(y)+\cdots\label{strdis}
\end{eqnarray}
In \cite{horo1} Horowitz and Wald proved that the conformal non-invariant (anomaly generating) part of the $T_{R}^{(1)ab}$ is local. The argument goes as follows. Consider the quantity $D^{ab}[g]=\langle 0_{in}|T^{ab}[g]|0_{in}\rangle-\langle 0_{out}|T^{ab}[g]|0_{out}\rangle$. Since $D^{ab}$ is the difference between two expectation values, the non-invariant regularization terms added in for them will be cancelled. Hence, $D^{ab}$ is conformal invariant, that is, it will transform with the naive conformal weight,
\begin{equation}
D^{ab}[\tilde{g}]=e^{6\omega}D^{ab}[g].
\end{equation}
From this equation it follows that
\begin{equation}
\langle 0_{in}|T^{ab}[\tilde{g}]|0_{in}\rangle-e^{6\omega}\langle 0_{in}|T^{ab}[g]|0_{in}\rangle
=\langle 0_{out}|T^{ab}[\tilde{g}]|0_{out}\rangle-e^{6\omega}\langle 0_{out}|T^{ab}[g]|0_{out}\rangle\label{local}
\end{equation}
Due to causality the left hand side of the above equation can only depend on quantities in the past lightcone, while the right hand side would only depend on those in the future lightcone. Therefore the difference $\langle 0_{in}|T^{ab}[\tilde{g}]|0_{in}\rangle-e^{6\omega}\langle 0_{in}|T^{ab}[g]|0_{in}\rangle$ can only involve local quantities.

Since the dissipation kernel comes from the nonlocal part of the stress energy tensor, the integral in Eq.~(\ref{strdis}) should transform as
\begin{eqnarray}
\int d^{4}y\sqrt{-\tilde{g}(y)}\,\tilde{H}_{A}^{abcd}(x,y)\tilde{h}_{cd}(y)
=e^{6\omega(x)}\int d^{4}y\sqrt{-g(y)}\,H_{A}^{abcd}(x,y)h_{cd}(y)
\end{eqnarray}
With $\tilde{h}_{ab}=e^{-2\omega}h_{ab}$, we have the conformal transformation for the dissipation kernel,
\begin{equation}
\tH_{A}^{abcd} (x,y) = e^{6 \om(x)}  H_{A}^{abcd} (x,y) e^{6 \om(y)}.\label{confdis}
\end{equation}
This is the same transformation rule as the noise kernel in Eq.~(\ref{confnoise}). We therefore see that the fluctuation-dissipation relation will be maintained under conformal transformation.

To be more specific we take the fluctuation-dissipation relation in the background spacetime $g_{ab}$ to be
\begin{equation}
N^{abcd}(x,x')=\int_{-\infty}^{\infty}dx'' K(x,x'')H_{A}^{abcd}(x'',x')
\end{equation}
where $K(x,x')$ is the fluctuation-dissipation kernel. From the conformal transformations of the noise and the dissipation kernels in Eqs.~(\ref{confnoise}) and (\ref{confdis}), we can readily obtain the fluctuation-dissipation relation in the conformally-related spacetime $\tilde{g}_{ab}$,
\begin{equation}
\tilde{N}^{abcd}(x,x')=\int_{-\infty}^{\infty}dx'' \tilde{K}(x,x'')\tilde{H}_{A}^{abcd}(x'',x'),
\end{equation}
where
\begin{equation}
\tilde{K}(x,x')=e^{6\omega(x)}K(x,x')e^{-6\omega(x')},
\end{equation}
is the transformation rule for the fluctuation-dissipation kernel.

\section{Conclusions and discussions}

We have obtained in this paper a semiclassical ELE, through conformal transformations from another one in a conformally-related spacetime. Under the conformal transformation, anomalous terms \cite{ottebrown,ottepage} will be appended to the influence action as the path integral measure is non-invariant. Using this modified influence action we are able to derive the conformal transformation of the in-in expectation value of the quantum stress energy tensor, which includes the term with the dissipation kernel. Together with the transformation of the noise kernel proven in \cite{EBRAH}, we have the transformation properties of all the quantities in the ELE. At the same time the fluctuation-dissipation relation under this transformation is attained.

This approach has been developed to facilitate further solution of the ELE from that of a simpler spacetime. An immediate application would be on conformally-flat spacetimes. For example, Martin and Verdaguer \cite{ver2} have given a detailed analysis on the Minkowski case. With their results our approach here will readily give the ELEs for all spatially-flat Friedmann-Robertson-Walker universes. When the conformal factor $\omega(\tau)\sim -{\rm ln}(-1/H\tau)$, one has the spatially-flat de Sitter universe which is relevant to issues in the inflationary cosmological model. Solutions to the ELE will then describe the behavior of metric fluctuations in the very early universe. (See \cite{enric} for recent progress in this direction.)

For the other conformally-flat spacetimes, like the closed and the open Friedmann universes, it is simpler to start with the Einstein and the open Einstein static spaces separately since their conformal vacua are different \cite{candow}. For the noise kernel this is the consideration in \cite{chohu2}. 
In addition to these cases the Rindler space will also be very interesting to look at. With the ELE in static spacetimes at hand, we shall be able to, using the approach in this paper, to obtain the ELE for all Friedmann universe, flat, closed or open. Subsequently, with the cosmological perturbation techniques (e.g.,  in \cite{kodsas}), it is possible to analyze the evolution of the cosmological perturbations around these background spacetimes with the quantum backreaction effects included.

\vspace{0.4in}

\end{document}